\documentstyle[prl,aps]{revtex}
\begin{document}
\title
{Kruskal Coordinates and Mass of Schwarzschild Black Holes}

\author{Abhas Mitra}
\address{Theoretical Physics Division, Bhabha Atomic Research Center,\\
Mumbai-400085, India\\ E-mail: amitra@apsara.barc.ernet.in}


\maketitle

\begin{abstract}
Schwarzschild coordinates ($r,t$) fail to describe the region within the
event horizon (EH), $(r\le r_g)$, of a Black Hole (BH) because the metric coefficients exhibit
singularity at $r =r_g$ and the radial geodesic of a particle appears to
be null ($ds^2 =0$) when actually {\em it must  be timelike} ($ds^2 >0$),
if $r_g >0$.  Thus, both the exterior and the interior regions of BHs
 are described by singularity free Kruskal coordinates.  However, we
{\em show} that, in this case too, $ds^2\rightarrow 0$ for $r \rightarrow
r_g$.  And this result can be physically reconciled only if the EH
 coincides with the central singularity or if the mass of
Schwarzschild black holes $M\equiv 0$.
\end{abstract}

\vskip 1cm
The concept of Black Holes (BHs) is one of the most important plinths of
modern physics and astrophysics. As is well known, the basic concept of BHs
actually arose more than two hundred years ago in the cradle of Newtonian
gravitation\cite{1}.
In General Theory of Relativity (GTR), the gravitational mass is less than
the baryonic mass ($M\le M_0$). Further, as the body contracts and emits
radiation $M$ keeps on decreasing progressively alongwith $r$. Thus, given
an initial gravitational mass $M_i$, one can not predict with certainty
 the value of $M_f$ when we would have $2M_f/r =1$ ($G=c=1$).
Neither are the values of $M_i$, $M_f$ and $M_0$  related by any
combination of fundamental constants though, it is generally
assumed that $M_i\approx M_f$. Ideally, one should solve the
Einstein equations analytically to fix the value of $M_f$ for a
given initial values of $M_i$ and $M_0$ for a realistic equation
of state (EOS) and energy transport properties. However even when
one does away with the EOS by assuming the matter to behave like
a dust, $p\equiv 0$, one does not obtain any unique solution if
the dust is inhomogeneous. Depending on the various initial
conditions and assumptions (like self-similarity) employed one
may end up finding either a BH or a ``naked
singularity''\cite{2}. By further assuming the dust to be {\em
homogeneous}
 Oppenheimer
and Snyder (OS)\cite{3} found  {\em asymptotic}
 solution of the  problem
by approximating Eq.(36) of their paper.
The region exterior to the event horizon ($r >r_g=2M$)
can be described by the Schwarzschild coordinates $r$ and $t$\cite{4,5}:
\begin{equation}
ds^2 = g_{tt} dt^2 +g_{rr} dr^2 + g_{\theta \theta} d\theta^2 + g_{\phi \phi} d\phi^2
\end{equation}
where
$g_{tt} =(1- 2M/r)$, $ g_{rr} =-(1-2M/r)^{-1}$, $ g_{\theta
\theta}=-r^2$, and $g_{\phi \phi} = -r^2 \sin^2 \theta$.
 Here, we are working with a spacetime
signature of +1, -1, -1, -1 and  $r$
has a distinct physical significance as the {\em invariant circumference radius}.
For $r>r_g=2M$, the worldline of a freeling falling radial material
particle is indeed timelike $ds^2 >0$  and the metric coefficients have the
right signature, $g_{tt} >0$, $g_{rr} <0$, $g_{\theta \theta} <0$ and
$g_{\phi \phi} <0$. But at $r=2M$, $g_{rr}$ blows
up and as $r <2M$, the $g_{tt}$ and $g_{rr}$ suddenly exchange their
signatures {\em though the signatures of $g_{\theta \theta}$ and $g_{\phi
\phi}$ remain unchanged}. This  is interpreted by saying that, inside the
event horizon, $r$ becomes ``time like'' and $t$ becomes
``spacelike''\cite{4,5}. However, we see that actually $r$
continues to retain, atleast partially, its spacelike character
by continuing to be ``invariant circumference radius''. Also,
note that, if physically measurable quantities like the Rimennian
curvature components behaved like $\sim M/r^3$ outside the EH,
they continue to behave in a similar manner, {\em and not like
$\sim M/t^3$} inside the EH. And it should be borne in mind here
that by a fresh relabelling or by any other means, the curvature
components can not be made to assume the form $\sim M/t^3$. One
particular reason for this is that, we would see later that,
inside the EH, we have $t=\infty$ while, of course, the value of
$r$ remains finite. Thus it may not actually be justified to
conclude that $r$ becomes the ``timelike coordinate'' inside the
EH even though $g_{rr}$ changes its sign. So far, it has not been
possible to resolve this enigma of the {\em duality} in the
behaviour of $r$ for $r <2M$, and the present paper intends to
attend to this problem. Since $ds$ is the proper time, we may
also write
\begin{equation}
ds^2=dt^2 \left( 1- {2M\over r} \right)
\end{equation}
Therefore, the radial geodesic of a {\em material particle} in the
Schwarzschild metric becomes,
 {\em unphysically} {\em null} ($ds^2=0$) and then
{\em spacelike} ($ds^2 <0$) as one moves inside the event horizon (EH).
In contrast, any physically meaningful coordinate system must be free of
such anomalies.
Although $g_{rr}$ blows up at $r=2M$, as mentioned before, the
curvature components of the Rimennian tensor behave perfectly normally at $r=r_g$,
$R^{ij}_{kl} \sim M/r^3$. Further,
the determinant of the metric coefficients continues to be
negative and finite $g = r^4 \sin^2 \theta~ g_{rr}~ g_{tt} = -r^4 \sin^2
\theta \le 0 $. Such realizations gave rise to the idea that the
Schwarzschild coordinate system suffers from a ``coordinate
singularity'' at the event horizon and must be replaced by some
other well behaved coordinate system. It is known that a comoving
coordinate system is naturally singularity free and Lemaitre
suggested that the region inside $r\le r_g$ may be represented by
such a coordinate system\cite{6} whereas the exterior region is
still described by the old Schwarzschild coordinates. It is only
in 1960 that Kruskal and Szekeres\cite{4,5,7} discovered a
one-piece coordinate system which can describe both the interior
and exterior regions of a BH. They achieved this by means of the
following coordinate transformation for the exterior region
(Sector I):
\begin{equation}
u=f_1(r) \cosh
{t\over 4M}; \qquad v=f_1(r) \sinh
{t\over 4M}; r\ge 2M
\end{equation}
where
\begin{equation}
f_1(r) = \left({r\over 2M} -1\right)^{1/2} e^{r/4M}
\end{equation}
It would be profitable to note that
\begin{equation}
{df_1\over dr} = {r\over 8M^2} \left({r\over 2M} -1\right)^{-1/2} e^{r/4M}
\end{equation}
And for the region interior to the horizon (Sector II), we have
\begin{equation}
u=f_2(r) \sinh
{t\over 4M};\qquad v=f_2(r) \cosh
{t\over 4M}; r\le 2M
\end{equation}
where
\begin{equation}
f_2(r) = \left(1- {r\over 2M}\right)^{1/2} e^{r/4M}
\end{equation}
and
\begin{equation}
{df_2\over dr} = {-r\over 8M^2} \left(1- {r\over 2M}\right)^{-1/2} e^{r/4M}
\end{equation}
Given our adopted signature of spacetime ($-2$), in terms of $u$ and $v$, the metric for the entire spacetime is
\begin{equation}
ds^2 = {32 M^3\over  r} e^{-r/2M} (dv^2 -d u^2) - r^2 (d\theta^2
+d\phi^2 \sin^2 \theta)
\end{equation}
The metric coefficients are regular everywhere except at the
intrinsic singularity $r=0$, as is expected. Note that, the
angular part of the metric remains unchanged by such
transformations and $r(u,v)$ continues to signal its intrinsic
spacelike nature. In either region we have
\begin{equation}
u^2-v^2= \left({r\over 2M} -1\right) e^{r/2M}
\end{equation}
so that
\begin{equation}
u^2-v^2 > 1; \qquad  u/v >\pm 1; \qquad r >2M,
\end{equation}
\begin{equation}
u^2-v^2 \rightarrow 0; \qquad u =\pm v; \qquad r= 2M
\end{equation}
and
\begin{equation}
u^2-v^2 <0; \qquad u/v < \pm 1; \qquad r <2M
\end{equation}
So, each of these above three  inequalities, and, in particular, the $r=0$ point
corresponds to not one but two conditions! 
\begin{equation}
v= \pm (1+u^2)^{1/2}
\end{equation}
Here, one point needs to be hardly overemphasized; astronomical
observations and experiments actually
conform to the idea that atleast far from massive bodies or probable BHs,
the spacetime is well described by the $r,t$ coordinate system.
In fact, although in the (normal) physical spacetime, in a spherically
symmetric spatial geometry (as defined by the implications of $r$ as an
``invariant circumference radius''), the physical singularity corresponds
to a mathematical point, in the Kruskal world view, this central
singularity corresponds to a pair of hyperbolas in the ($u-v$) plane.
While the ``+ve'' sign of equation corresponds to the central BH
singularity, the ``-ve'' sign corresponds to the singularity inside a
so-called {\em White Hole} which may spew out mass-energy spontaneously in
``our universe''\cite{4,5}. The white hole singularity belongs to ``other universe''
whose presence is suggested by the fact that the Kruskal metric remains
unaffected by the following additional transformations:
\begin{equation}
u=-f_1(r) \cosh
{t\over 4M};\qquad v=-f_1(r) \sinh
{t\over 4M}; r\ge 2M
\end{equation}
defining Sector (III) and
\begin{equation}
u=-f_2(r) \sinh
{t\over 4M}; \qquad v=-f_2(r) \cosh
{t\over 4M}; r\le 2M
\end{equation}
defining Sector (IV). Thus not only does the region interior to the EH
correspond to two different universes, (Sector II and IV) but the
structure of the physical spacetime outside the EH,
 too, effectively corresponds to two universes (Sector I and
III).
If there exists $N$ number of BHs, the (normal) physical spacetime may be
much more complex.
  The aim of this paper is
to {\em explicitly verify} whether the (radial) geodesics of
material particles are indeed {\em timelike} at the EH which they
must be if this idea of a finite mass Schwarzschild BH is
physically correct. First we focus attention on the region $r\ge
2M$ and differentiate Eq.(3) to see
\begin{equation}
{du\over dr} = {\partial u\over \partial r} + {\partial u\over \partial t}
{dt\over dr}= {df\over dr} \cosh {t\over 4M} + {f\over 4M} \sinh
{t\over 4M}{dt\over dr}
\end{equation}
Now by using Eq. (4-6) in the above equation, we find that
\begin{equation}
{du\over dr} = {ru\over 8M^2} (r/2M -1)^{-1} + {v\over 4M} {dt\over dr}; \qquad
r\ge 2M
\end{equation}
and
\begin{equation}
{dv\over dr} = {rv\over 8M^2} (r/2M -1)^{-1} + {u\over 4M} {dt\over dr};
\qquad r\ge 2M
\end{equation}
By dividing equation (18) by (19), we obtain
\begin{equation}
{du\over dv} = {{ru\over 2M}  + v {dt\over dr} (r/2M-1) \over
{rv\over 2M}  + u {dt\over dr} (r/2M -1)}
\end{equation}
Similarly, starting from Eq. (6), we end up obtaining a form of
$du/dv$ for the region $r <2M$ which is exactly similar to the
foregoing equation. Now, by using Eq.(12) ($u=\pm v$) in
 Eq. (20), we promptly find that
\begin{equation}
{du\over dv}\rightarrow {{\pm r\over 2M}  +  {dt\over dr} (r/2M-1) \over
{ r\over 2M}  \pm {dt\over dr} (r/2M -1)} \rightarrow \pm 1;\qquad r\rightarrow 2M
\end{equation}
Thus, we are able to find the precise value of $du/dv$  at the EH in a most general
manner
{\em irrespective of the precise relationship} between $t$ and $r$.
Armed with this value of $du/dv$, we are in a position now to complete our
task by
 rewriting the {\em radial part} of the Kruskal metric ($d\theta
=d\phi =0$) as
\begin{equation}
ds^2 = {32 M^3\over  r} e^{-r/2M} dv^2 \left[1- \left({du\over dv}\right)^2\right]
\end{equation}
Or,
\begin{equation}
ds^2 = 16 M^2 e^{-1} dv^2 (1-1)=0; \qquad r=2M
\end{equation}
We have found that {\em for the Lemaitre
coordinate too},  $ds^2 =0$ at $r=2M$.
This implies that although the metric coefficients can be made to appear
 regular, the radial
geodesic of a {\em material particle becomes null} at the event
horizon of a finite mass BH in contravention of the basic
premises of GTR! And since, now, we can not blame the coordinate
system to be faulty for this occurrence, the only way we can
explain this result is that {\em the Event Horizon itself
corresponds to the physical singularity} or, in other words, the
mass of the Schwarzschild BHS $M\equiv 0$. And then, the entire
conundrum of ``Schwarzschild singularity'', ``swapping of spatial
and temporal characters by $r$ and $t$ inside the event horizon
({\em when the angular part of all metrics suggest that $r$ has a
spacelike character even within the horizon}), ``White Holes''
and ``Other Universes'' get resolved. Here we recall the
conjecture of Rosen\cite{8} ``so that in this region $r$ is
timelike and $t$ is spacelike. However, this is an impossible
situation, for we have seen that $r$ defined in terms of the
circumference of a circle so that $r$ is spacelike, and we are
therefore faced with a contradiction. We must conclude that the
portion  of space corresponding to $r< 2M$ is
non-physical. This is a situation which a coordinate
transformation even one which removes a singularity can not
change. What it means is that the surface $r=2M$ represents the
boundary of physical space and should be regarded as an
impenetrable barrier for particles and light rays.'' This idea of
Rosen is also in accordance with the idea of Einstein that the
Schwarzschild type singularity is unphysical and can not occur
for realistic cases\cite{9}.
 And this paper indeed shows that {\em in order that the radial worldlines
of free falling material particles do not become null at a mere
coordinate singularity}, Nature (GTR) refuses to have any
spacetime within the EH.

Although, having made our basic point, {\em we could have ended
this paper at this point}, for the sake of further insight, we
shall study the behaviour of $ds^2$
 for the entire spacetime by, again assuming, for a moment, the existence of a finite mass BH.
It can be found  that in the region $r>2M$, one would indeed have $ds^2 >0$ for $r >2M$.
And to see the
behaviour of $du/dv$ inside the EH, we recall the relationship between $t$
and $r$ (see pp. 824 of ref.[4] or pp. 343 of ref.[5]):
\begin{equation}
{t\over 2M} = \ln\mid{(r_\infty/2M-1)^{1/2} + \tan{(\eta/2)} \over (r_\infty
/2M-1)^{1/2} - \tan{(\eta/2)}}\mid + 2M\left({r_\infty\over 2M}-1\right)^{1/2}
\left[\eta + \left({r_\infty\over 4M}\right)(\eta +\sin \eta)\right]
\end{equation}
where the particle is released with zero velocity from $r=r_\infty$ at $t=0$
and the ``cyclic'' coordinate $\eta$ is defined by
\begin{equation}
r ={r_\infty\over 2} (1+\cos \eta)
\end{equation}
Since
$\tan{(\eta/2)} = (r_\infty/r -1)$
we find from Eq. (24) that, as $r\rightarrow 2M$, the logarithmic
term blows up and $t\rightarrow \infty$, which is a well known
result. And since $t$ continues to increase as the particle
enters the EH, we have the general result that $t =\infty$ for
$r\le 2M$. In this limit, we have
\begin{equation}
\cosh{t\over 4M} \rightarrow \sinh{t\over 4M} \rightarrow {e^{t/4M}\over 2} =\infty
\end{equation}
 Consequently, even though,
$u^2 -v^2$ continues to be finite  we obtain
\begin{equation}
{u\over v} =\pm 1; \qquad r\le 2M
\end{equation}
Hence we obtain a more general form of Eq. (21)
\begin{equation}
{du\over dv}\rightarrow
 \pm 1;\qquad r\le 2M
\end{equation}
irrespective of the {\em precise} form of $dt/dr$. Then from Eq.
(22), we find that the metric would {\em continue to be null} for
$r <2M$:
\begin{equation}
ds^2 =0; \qquad r \le 2M
\end{equation}
 And this unphysical happening is of course
avoided when we realize that $M=0$ and there is no additional spacetime
between the EH and the central singularity.
We may mention now that we have recently shown that the OS work too
actually suggests that the mass of the resultant BH must be $M\equiv 0$\cite{10}.
The basic reason for this assertion is extremely simple. The Eq.(36) of
OS paper connects $t$ and $r$ through a relationship which, for large
values of $t$ is
\begin{equation}
t \sim \ln{y^{1/2} +1\over y^{1/2} -1}
\end{equation}
where at the boundary of the fluid
\begin{equation}
y={r\over r_g} = {r\over 2M}
\end{equation}
Since the argument of a {\em logarithmic function can not be negative}, in
order that $t$ is definable at all
 that we must have
\begin{equation}
y={r\over 2M} \ge 1; \qquad {2M\over r}\le 1
\end{equation}
Thus atleast for the collapse of a homogeneous dust, ``trapped
surfaces'' do not form and if the collapse continues to the point
$r\rightarrow 0$ we must have $M_f\rightarrow 0$. This
independent finding is in complete agreement with what we have
shown in the present paper that Schwarzschild BHs must have
$M=0$. Although, there is no modulus here in the argument of the
logarithmic of Eq. (30) (unlike Eq. [24]), some readers may wish
there were one. Even if one imagined the existence of such a
modulus, one would run into contradiction in the following way.
Of course we will have $t \rightarrow \infty$ as $r\rightarrow
2M$. But during the collapse if one would enter $r <2M$ (if $M
>0$), $t$ {\bf would start decreasing}!

However, unlike the case of Newtonian gravity, in GTR, $M=0$
state need not correspond to a configuration with zero baryonic
mass. The $M=0$ state is simply one in which the negative
gravitational energy exactly offsets the positive energy
associated with $M_0$ and internal energy, and may indeed
represent a physical singularity with infinite energy density and
tidal acceleration. For instance, if the collapse process leads
to the $y=1$ limit, then the curvature components $R^{ij}_{kl}
\sim M/ r^3 \sim r^{-2} \rightarrow \infty$ as $r\rightarrow 0$.
Note also that, the metric coefficients $g_{uu}$ and $g_{vv}$ for
the zero-mass BH blow up in a similar fashion at the EH. It may
be noted that the ``naked singularities'' too may be
characterized by $M=0$\cite{11}. In the context of the dust
collapse, we see that, for, $M=0$, the proper time for the
formation of the BH would be infinite
\begin{equation}
\tau = \pi \left({r_\infty^3\over 8 M}\right)^{1/2} =\infty
\end{equation}
Further, we have shown elsewhere that the crucial condition (32),
$y\ge 1$, is valid not only for the OS problem, but also for any
generic spherical gravitational collapse\cite{12}. And similarly,
$\tau\rightarrow
\infty$ as $r\rightarrow 0$ not only for dust collapse, but also
for the collapse of any physical fluid\cite{12}. Thus at any
given finite proper time there would be no BH, and on the other
hand there could be dynamically collapsing configurations with
arbitrary high surface redshifts. In fact it can be found that
the {\em proper length} of a radial geodesic becomes infinite
too\cite{12}. And therefore, even if, such dynamically
configurations with large surface red-shifts may be collapsing
with relativistic velocities,
 the collapse process
will never terminate in any finite amount of time. This happens
because spacetime would get infinitely stretched by infinite
curvature near $r=0$. This is a purely general relativistic
effect, and is difficult to comprehend by ``common astronomical
sense''. Observationally, such configurations may be identified
as Black Holes. And if some of these configurations are
collapsing with nearly free fall speed, accretion onto such
configurations
 would emit little radiation if the accretion flow happens to
 be advection dominated.
To conclude, irrespective of the observational consequences, we
have {\em directly} shown that, if GTR is correct, Schwarzschild
BHs must have $M\equiv 0$ in order that the radial geodesics of
material particles remain timelike at a finite value of $r$.

\end{document}